\documentstyle[11pt,moriond,epsfig]{article}

\begin{document}

\title{COHESION, CONDUCTANCE, AND CHARGING EFFECTS IN A\\METALLIC
NANOCONTACT
\footnote{Published in {\em Quantum Physics at Mesoscopic Scale}, 
D.\ C.\ Glattli, M.\ Sanquer and J.\ Tran Thanh Van eds.\
(EDP Sciences, Les Ulis, France, 2000), pp. 49-53.}
}

\author{C.\ A.\ Stafford,$^1$
F.\ Kassubek,$^{1,2}$ 
J.\ B\"urki,$^{1,3,4}$
Hermann Grabert,$^2$ and
D.\ Baeriswyl$^3$}

\address{$^1$Department of Physics, University of Arizona, Tucson, AZ 85721\\
$^2$Fakult\"at f\"ur Physik, Albert-Ludwigs-Universit\"at,
D-79104 Freiburg, Germany\\
$^3$Institut de Physique Th\'{e}orique, Universit\'{e} de Fribourg,
CH-1700 Fribourg, Switzerland\\
$^4$Institut Romand de Recherche Num\'erique en 
Physique des Mat\'eriaux, CH-1015 Lausanne, Switzerland}

\maketitle\abstracts{ 
The conducting and thermodynamic properties of ballistic
metallic nanocontacts with smooth shapes are investigated.  All properties
are related to the electronic scattering matrix, which is evaluated in the
WKB approximation for independent electrons and in the self-consistent
Hartree approximation for interacting electrons.
Mesoscopic oscillations of order 1nN in the cohesive force 
and of order $e$ in the contact charge
are predicted when a metallic nanocontact is pulled apart,
which are synchronized with quantized jumps in the conductance.}

\section{Introduction}
\label{sec.intro}
 
In a seminal experiment published in 1996, Rubio, Agra\"{\i}t, 
and Vieira \cite{nanoforce} measured simultaneously the electrical conductance
and cohesive force of Au--Au contacts with diameters ranging from several
\AA ngstroms to several nanometers.  They observed oscillations
in the cohesive force of order 1nN as the contact was pulled apart,
which were synchronized with jumps of order $G_0=2e^2/h$ in the conductance
$G$.
Similar results were obtained independently
by Stalder and D\"urig.\cite{nanoforce2} 
In a previous article,\cite{sbb} it was argued that the mesoscopic force
oscillations, like the corresponding conductance steps, 
can be understood by considering the nanocontact as a waveguide for the 
conduction electrons (which are responsible for both conduction and cohesion in
simple metals): each quantized mode transmitted through the contact 
contributes $2e^2/h$ to the conductance,\cite{beenakker} 
and a force of order $E_F/\lambda_F$ to the cohesion,\cite{sbb} where
$\lambda_F$ is the de Broglie wavelength at the Fermi energy $E_F$.
Recent experiments \cite{yanson} suggest that mode 
quantization plays an important role in the cohesion of Na nanocontacts 
for $G \leq 120 G_0$.
In Ref.\ 3 and several subsequent 
works,\cite{comment.zwerger}$^-$\nocite{wnw,zwerger2}\cite{jerome}
an independent-electron model was used to describe the cohesion
and conductance of metallic nanocontacts within the grand canonical ensemble.
Recently, several attempts have been made to include the effects of
electron-electron interactions in the extended
Thomas-Fermi approximation \cite{landman} and in the local density 
approximation; \cite{lda} however, these calculations have 
utilized the canonical ensemble, which is not appropriate for an open 
mesoscopic system, such as a metallic nanocontact.  

In this article, we show how
to treat electron-electron interactions self-consistently in open mesoscopic
systems via a generalization of the scattering-matrix approach 
of Ref.\ 3.  
In Sec.\ \ref{sec.free}, 
we describe the scattering-matrix formalism for independent
electrons, and show that the bare mesoscopic charge oscillations induced by the 
quantum confinement are small, of order $e$ in nanometer scale contacts.
In Sec.\ \ref{sec.interaction}, we derive the self-consistent Hartree 
approximation using the electronic scattering matrix, and obtain integral
equations for the screened charge and grand canonical potential within 
linear response.  The Hartree
equations are solved for the case of a discrete-potential
model, and it is shown that the Hartree correction to the cohesive force
is small, justifying the use of the independent-electron model.
\cite{sbb,comment.zwerger}$^-$\nocite{wnw,zwerger2}\cite{jerome}

\section{Independent electron model of a metallic nanocontact}
\label{sec.free}

We first model the metallic nanocontact as a system of
noninteracting electrons confined along the $z$-axis 
by a potential $U_0({\bf x})$ with a saddle-point at ${\bf x}=0$, 
and assume $U_0$ varies slowly, i.e.,
$\partial U_0/\partial z\ll E_F/\lambda_F$ 
$\forall \,{\bf x}=(x,y,z)$.  Neglecting small terms of order
$\partial U_0/\partial z$,
the Schr\"odinger equation is separable into a series of one-dimensional
scattering problems with effective 
potential barriers $\varepsilon_\nu(z)$, 
the eigenvalues of the two-dimensional Schr\"odinger equation in 
planes of constant $z$.  The scattering matrix is then diagonal, and for
a system with inversion symmetry about $z=0$, it is 
found in an extended WKB approximation to be \cite{sbb}
\begin{equation}
S_{\mu\nu}(E) = \delta_{\mu\nu}\, e^{i\Theta_\nu (E)}
\left(\begin{array}{cc} i R_\nu^{1/2}(E) & T_\nu^{1/2}(E) \\
T_\nu^{1/2}(E) & i R_\nu^{1/2}(E) \end{array}\right),
\label{s.wkb}
\end{equation}
where
\begin{equation}
\Theta_\nu(E) = (2m/\hbar^2)^{1/2} \int_{\varepsilon_\nu(z)<E} dz \, 
[E - \varepsilon_\nu(z)]^{1/2}
\label{phase.wkb}
\end{equation}
is the scattering phase shift, and $R_\nu(E)$ and $T_\nu(E)$ are the reflection
and transmission probabilities for mode $\nu$, which satisfy
$R_\nu(E) + T_\nu(E)=1$. The transmission probability for mode $\nu$ may be
expressed \cite{cond.wkb} as
$T^{-1}_\nu(E)=1+{\cal T}_\nu(E)^{-1}$, where
\begin{equation}
\ln {\cal T}_\nu(E) =
\left\{\begin{array}{cc} 
{\displaystyle 
-2(2m/\hbar^2)^{1/2}\int_{\varepsilon_\nu(z)>E} dz\,
[\varepsilon_\nu(z) -E ]^{1/2}}, & E < \varepsilon_\nu(0),\\
\mbox{ } & \\
{\displaystyle 
2(2m/\hbar^2)^{1/2}\int_{\varepsilon_\nu(z)>2\varepsilon_\nu(0)
-E} dz\,
[\varepsilon_\nu(z) +E -2\varepsilon_\nu(0)]^{1/2}}, & E > \varepsilon_\nu(0).
\end{array}\right.
\label{t.wkb}
\end{equation}

Given the scattering matrix (\ref{s.wkb})--(\ref{t.wkb}), 
the electrical conductance of
the nanocontact may be obtained from the Landauer formula,\cite{beenakker}
and all equilibrium thermodynamic properties of the system may be obtained
from the density of states $g(E)$,
\begin{equation}
G=\frac{ 2 e^2}{h} \sum_\nu T_\nu(E_F),\;\;\;\;\;\;
g(E)=\frac{1}{2\pi i} \mbox{Tr}\left\{
S^{\dagger} \frac{\partial S}{\partial E}
- S \frac{\partial S^{\dagger}}{\partial E}
\right\}
= \frac{2}{\pi} \sum_\nu \frac{\partial \Theta_\nu}{\partial E}.
\label{Gandos.wkb}
\end{equation}
The grand canonical potential, which governs the energetics of the 
nanocontact is \cite{sbb}
\begin{equation}
\Omega=\int_0^{E_F} dE \, g(E) (E-E_F)
=-\frac{8E_F}{3\lambda_F} 
\sum_\nu \int_{\varepsilon_\nu(z) < E_F}
dz\,\left(1-\frac{\varepsilon_\nu(z)}{E_F}\right)^{3/2}.
\label{e0.wkb}
\end{equation}
The total number of electrons in the nanocontact is 
\begin{equation}
N_-=\int_0^{E_F}dE\, g(E)
=\frac{4}{\lambda_F} \sum_\nu \int_{\varepsilon_\nu(z) < E_F}
dz\,\left(1-\frac{\varepsilon_\nu(z)}{E_F}\right)^{1/2}.
\label{q.wkb}
\end{equation}
Eqs.\ (\ref{Gandos.wkb}), (\ref{e0.wkb}), and (\ref{q.wkb}) are given at zero
temperature for simplicity.
When the nanocontact is elongated, the contact area decreases, steepening
the saddle in the confining potential, and hence increasing 
the mode energies $\varepsilon_\nu(z)$.
The cohesive force is given by $F=-\partial \Omega/\partial L$.

\subsection{Hard-wall confinement}
\label{sec.hardwall}

We now specialize to the case where electrons are confined within the 
nanocontact by impenetrable walls, with $U_0=0$  on the interior.
In this case, the condition for the validity of
Eqs.\ (\ref{e0.wkb}) and (\ref{q.wkb}) is $|dD(z)/dz|\ll 1$, where
$D(z)$ is the diameter of the contact.\cite{sbb}
In order to separate the macroscopic contributions to $\Omega$ and $N_-$
from the quantum fluctuations, it is useful \cite{comment.zwerger} to express
the sum over transverse modes $\nu$ in 
Eqs.\ (\ref{e0.wkb}) and (\ref{q.wkb}) in terms of an integral over a local
transverse density of states 
$\sum_\nu\int dz\,(\cdots) \rightarrow \int dE \int dz
\, \partial N_\perp(E,z)/ \partial E \, (\cdots)$. 
The total number of transverse modes 
$N_\perp(E,z)$ with $\varepsilon_\nu(z) < E$ may be determined with the
aid of the Weyl expansion \cite{comment.zwerger}
\begin{equation}
N_\perp(E,z) = \frac{k_E^2 {\cal A}(z)}{4\pi} - \frac{k_E \partial {\cal A}
(z)}{4\pi}
+ \frac{1}{6} + \delta N_\perp(E,z),
\label{n.perp}
\end{equation}
where ${\cal A}(z)$ and $\partial {\cal A}(z)$ are the cross-sectional area,
and circumference, respectively, of the contact as function of $z$, and
$\delta N_\perp(E,z)$ is a fluctuating correction determined by
the spectrum of quantized transverse modes, whose average is zero.
Using Eq.\ (\ref{n.perp}), Eqs.\ (\ref{e0.wkb}) and (\ref{q.wkb}) may be
rewritten as
\begin{equation}
\frac{\Omega}{E_F}  =  
-\frac{2k_F^3 {\cal V}}{15\pi^2} + \frac{k_F^2 {\cal S}}{16\pi} -
\frac{2k_F L}{9\pi} + \frac{\delta \Omega}{E_F},\;\;\;\;\;\;
N_-  =  \frac{k_F^3 {\cal V}}{3\pi^2} - \frac{k_F^2 {\cal S}}{8 \pi}
+ \frac{k_F L}{3\pi} + \delta N_-,
\label{eandq.weyl}
\end{equation}
where ${\cal V}$ is the volume of the nanocontact, ${\cal S}$ 
is its surface area, 
$L$ is its length, and $\delta \Omega$ and $\delta N_-$ are oscillatory
quantum corrections whose average is zero.

During elongation, it is assumed that the positive background charge density
remains constant, i.e., that the system is incompressible.\cite{wnw}
Hard-wall boundary conditions impose a node of the electron wavefunction
at the boundary, and lead to a depletion of electrons within a distance
of order $\lambda_F$ from the boundary [second term on the right of Eq.\
(\ref{eandq.weyl})].  To reconcile the constraint of incompressibility
and the Dirichlet boundary conditions, we assume that the shape $D(z)$ of
the nanocontact evolves during elongation in such a way that the macroscopic
part $\bar{N}_-$ of the electron number of the system 
[first three terms on the right of Eq.\ (\ref{eandq.weyl})] 
remains constant,\cite{constraint} and is neutralized by an equal and 
opposite positive background charge.  The net charge imbalance on the 
nanocontact (neglecting screening) is thus $\delta Q_0=-e\delta N_-$,
which we show to be small. 
Differentiating $\Omega$ with respect to $L$ with the constraint
$\bar{N}_-=\mbox{const.}$, and using Eq.\ (\ref{eandq.weyl}), one finds
\begin{equation}
F=-\left. \frac{\partial \Omega}{\partial L}\right|_{\bar{N}_-}
= -\frac{\sigma_{\cal V}}{5} \frac{\partial {\cal S}}{\partial L}
+ \frac{2}{5}\Delta F_{\rm top} + \delta F,
\label{force.weyl}
\end{equation}
where $\sigma_{\cal V}=E_F k_F^2/16\pi$ 
is the surface energy \cite{sbb} of a noninteracting electron gas
at fixed ${\cal V}$ and $\Delta F_{\rm top} =4E_F/9\lambda_F$
is the mean mesoscopic suppression \cite{comment.zwerger}
of the cohesive force at fixed ${\cal V}$.  
The reduction of the surface energy by
a factor of 5 has been discussed by Lang.\cite{lang}
Importantly, since the constraint $\bar{N}_-=\mbox{const.}$ differs from
the constraint ${\cal V}=\mbox{const.}$ only by terms of order 
$(k_F D_{\rm min})^{-1}$, the mesoscopic fluctuations $\delta F$ and 
$\delta N_-$ are quite insensitive to the choice of constraint.

\begin{figure}
\hspace*{-7.5mm}
\epsfig{figure=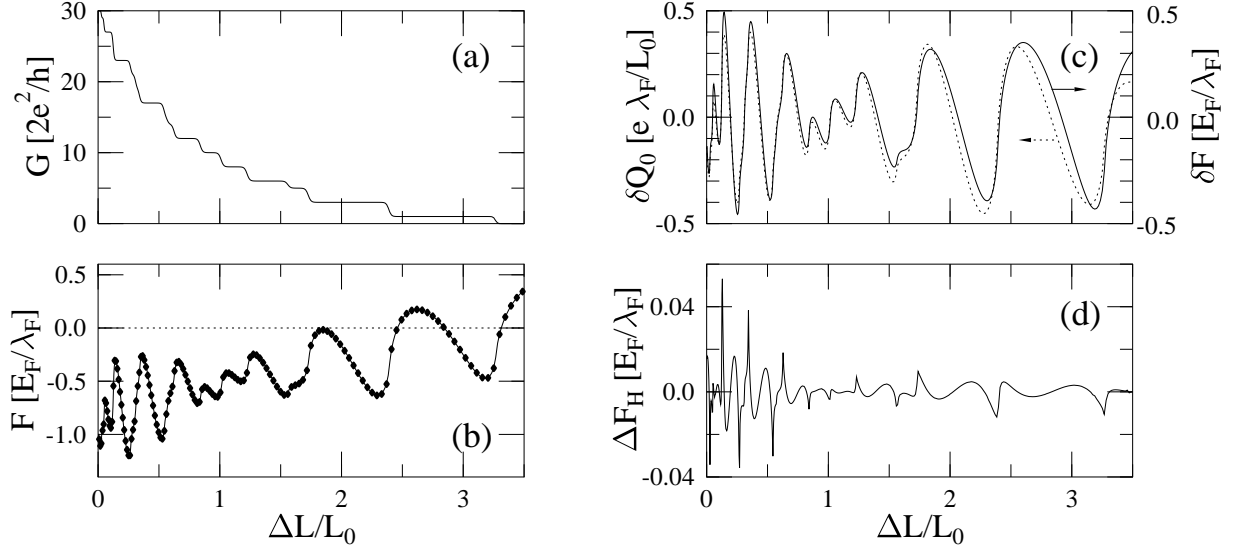,
width=17.6cm}
\caption{Properties of an axially symmetric nanocontact 
with diameter $D(z)=D_{\rm min}+(D_{\rm max}-D_{\rm min})(2z/L)^2$ as a
function of the relative elongation $\Delta L/L_0$, where $k_F D_{\rm max}
=24$ and $k_F L_0=40$:
(a) Electrical conductance $G$; (b) cohesive force $F$ in the
independent-electron model (solid curve) and in the self-consistent 
Hartree approximation (points); (c) mesoscopic charge oscillations
$\delta Q_0$ 
(neglecting screening) and force oscillations $\delta F$; 
(d) self-consistent
Hartree correction to the cohesive force in the constant interaction
model [Eq.\ (\ref{result.cim})].
}
\label{fig.1}
\end{figure}

Figure \ref{fig.1} shows the electrical 
conductance $G$ of a metallic nanocontact, calculated from Eqs.\
(\ref{t.wkb}) and (\ref{Gandos.wkb}), 
and its cohesive force $F$, calculated from
Eqs.\ (\ref{e0.wkb}) and (\ref{force.weyl}).  The conductance decreases in
steps of $2e^2/h$ and $4e^2/h$ (depending on whether the transverse mode 
involved is singly or doubly degenerate) as the contact is elongated, and
the force exhibits oscillations of amplitude $\sim 
E_F/\lambda_F$, which are synchronized with the conductance steps.
The overall magnitude of the force is reduced relative to that obtained in
a previous calculation \cite{sbb}
using a constraint of constant volume, but the force
oscillations $\delta F$ are {\em virtually identical} to those obtained in
Ref.\ 3.  Noting that $E_F/\lambda_F\simeq 0.76\mbox{nN}$ for Na,
one finds that the force shown in 
Fig.\ \ref{fig.1}(b) is also virtually identical to
that calculated for a Na nanowire 
by Yannouleas et al.,\cite{landman} suggesting that the more realistic jellium
confinement potential they utilized does not significantly
modify the cohesion.  

The net charge $\delta Q_0$ on the nanowire
(neglecting screening) is also shown in Fig.\ \ref{fig.1}.
Interestingly, we find that $\delta Q_0/L_0 \simeq e \delta F/E_F$,  
where $L_0$ is the initial length of the (initially cylindrical) nanocontact.
This result
is consistent with the result of Kassubek et al.\cite{wnw} for an abrupt
geometry, and will be discussed in more detail elsewhere.  Here it suffices to
note that since $|\delta Q_0|\ll e\bar{N}_-$, 
the electron-electron interactions in
the nanowire may be treated using linear-response theory.

\section{Interacting electron model}
\label{sec.interaction}

\subsection{Self-consistent Hartree approximation}
\label{sec.hartree}

In an interacting model, the confinement potential of the nanocontact
is determined self-consistently via the Coulomb interaction: 
$U_0({\bf x})\rightarrow U({\bf x})= 
\int d^3 y \; V({\bf x}-{\bf y})\delta\rho({\bf y})$
(Hartree approximation), where
$\delta\rho({\bf x}) \equiv \rho_-({\bf x}) + \rho_+({\bf x})$
is the net charge density, $V({\bf x})$ is the Coulomb interaction, and 
$\rho_+({\bf x})$ describes the fixed positive background.
The grand canonical potential becomes
\begin{equation}
\Omega = \Omega_0[U({\bf x})] - \frac{1}{2}\int d^3 x
\left[\rho_{-}({\bf x}) - \rho_{+}({\bf x})\right]U({\bf x}),
\label{e.hartree}
\end{equation}
where $\Omega_0[U({\bf x})]$ is given by Eq.\ (\ref{e0.wkb}) with 
$U_0({\bf x})$ replaced by $U({\bf x})$.  
The second term in Eq.\ (\ref{e.hartree}) corrects for the double counting
of electron-electron interactions in $\Omega_0[U({\bf x})]$.
The charge density of the conduction electrons is
\begin{equation}
\rho_-({\bf x}) =\frac{\delta \Omega_0}{\delta U({\bf x})} =
-\frac{1}{2\pi i}\int_0^{E_F} dE\, 
\mbox{Tr} \left\{S^{\dagger} \frac{\delta S}{
\delta U({\bf x})} - S \frac{\delta S^{\dagger}}{\delta U({\bf x})}\right\}.
\label{rho.hartree}
\end{equation}

\subsection{Linear response}
\label{sec.hartree.linear}

Motivated by the smallness of the bare charge fluctuations 
$\delta Q_0 =\int d^3x\,\delta \rho_0({\bf x})$,
we expand Eq.\ (\ref{rho.hartree}) to linear order in $U({\bf x})$, obtaining
a self-consistent equation for the screened charge density 
\begin{equation}
\int d^3 y \; \epsilon({\bf x},{\bf y}) \delta \rho({\bf y}) =
\delta\rho_0({\bf x}), \;\;\;\;\;\;
\epsilon({\bf x},{\bf y}) = \delta({\bf x}-{\bf y}) -
\int d^3 z \, \left.\frac{\delta \rho({\bf x})}{\delta U({\bf z})}
\right|_{U=0} \!\!\!\!\!\!  V({\bf z}-{\bf y}),
\label{eq.screen}
\end{equation}
where $\epsilon({\bf x},{\bf y})$ is the nonlocal Hartree dielectric 
function and
\begin{equation}
\frac{\delta \rho({\bf x})}{\delta U({\bf y})} = -\frac{1}{2\pi i}
\int_0^{E_F} dE\, \mbox{Tr} 
\left\{\frac{\delta S^{\dagger}}{\delta U({\bf y})}
\frac{\delta S}{\delta U({\bf x})} + S^{\dagger} \frac{\delta^2 S}{
\delta U({\bf y})\delta U({\bf x})} - \mbox{H.c.}\right\}.
\label{drhodU}
\end{equation}
Expanding Eq.\ (\ref{e.hartree}) in powers of $U({\bf x})$
and using Eq.\ (\ref{eq.screen}), one finds
\begin{equation}
\Omega = \Omega_0[0] + \frac{1}{2} \int d^3 x \int d^3 y \;\;
\delta \rho_0 ({\bf x}) \tilde{V}({\bf x},{\bf y}) \delta \rho_0({\bf y})
+ {\cal O}(\delta \rho_0^3),
\label{e.linres}
\end{equation}
where the screened interaction is given by
$\tilde{V}({\bf x},{\bf y}) = \int d^3 z \; V({\bf x}-{\bf z}) \epsilon^{-1}
({\bf z},{\bf y})$
and the inverse dielectric function $\epsilon^{-1}$ satisfies
$\int d^3 z \; \epsilon^{-1}({\bf x},{\bf z})\epsilon({\bf z},{\bf y}) =
\int d^3 z \; \epsilon({\bf x},{\bf z})\epsilon^{-1}({\bf z},{\bf y}) =
\delta({\bf x}-{\bf y})$.  Eq.\ (\ref{e.linres}) is a familiar result of
linear-response theory; the new element here is that the dielectric function is
calculated from the scattering matrix [Eqs.\ (\ref{eq.screen}) and 
(\ref{drhodU})] and is nonlocal due to quantum confinement.

\subsection{Constant interaction model}
\label{sec.cim}

A simple case for which Eqs.\ (\ref{eq.screen}) and (\ref{e.linres})
can be solved easily is the so-called constant interaction model,
$V({\bf x},{\bf y})=C^{-1}$ if ${\bf x}, {\bf y} \in {\cal V}$ and 
0 otherwise.  $C$ plays the role of the capacitance of the nanocontact
to its environment, and Eqs.\ (\ref{eq.screen}) and (\ref{e.linres})
become:
\begin{equation}
\delta Q = \frac{\delta Q_0}{1+ e^2 g(E_F)/C},\;\;\;\;\;\;\;\;\;
\Omega = \Omega_0[0] + \frac{1}{2} \frac{\delta Q_0^2}{
C+e^2 g(E_F)} + {\cal O}(\delta Q_0^3).
\label{result.cim}
\end{equation}
The result (\ref{result.cim}) confirms a previous conjecture \cite{wnw} based
on simple physical arguments.  An upper bound on the magnitude of
$\Delta \Omega_H = \Omega - \Omega_0[0]$ is obtained by setting $C=0$, which
enforces global 
charge neutrality.  The Hartree correction to the cohesive force
$\Delta F_H = -\partial \Delta \Omega_H/\partial L$ is plotted in Fig.\
\ref{fig.1} for $C=0$.  $\Delta F_H$ is over an order of magnitude smaller
than $\delta F$, and gives an indiscernible correction to the total force,
suggesting
that the independent electron model of metallic nanocohesion
\cite{sbb,comment.zwerger}$^-$\nocite{wnw,zwerger2}\cite{jerome}
is well justified.

\section*{Acknowledgments}

This work was supported in part by the Deutsche 
Forschungsgemeinschaft through grant SFB 276 and by Swiss National Foundation
Grant No.\ 4036-044033.


\section*{References}
\begin{thebibliography}{99}


\bibitem{nanoforce} C. Rubio, N. Agra\"{\i}t, and S. Vieira, Phys. Rev. Lett.
{\bf 76}, 2302 (1996). 

\bibitem{nanoforce2} A. Stalder and U. D\"urig, Appl. Phys. Lett. {\bf 68},
637 (1996). 



\bibitem{sbb} C. A. Stafford, D. Baeriswyl, and J. B\"urki, Phys. Rev. Lett.
{\bf 79}, 2863 (1997).

\bibitem{beenakker} For a review, see
C. W. J. Beenakker and H. van Houten, 
Solid State Phys. {\bf 44}, 1 (1991).

\bibitem{yanson} A. I. Yanson, I. K. Yanson, and J. M. van Ruitenbeek,
Nature {\bf 400}, 144 (1999).

\bibitem{comment.zwerger} C. H\"oppler and W. Zwerger, Phys. Rev. Lett.
{\bf 80}, 1792 (1998).

\bibitem{wnw} F. Kassubek, C. A. Stafford, and H. Grabert, Phys. Rev.  B
{\bf 59}, 7560 (1999).

\bibitem{zwerger2} C. H\"oppler and W. Zwerger, Phys. Rev. B {\bf 59},
R7849 (1999).

\bibitem{jerome} J. B\"urki, C. A. Stafford, X. Zotos, and 
D. Baeriswyl, Phys.\ Rev.\ B {\bf 60}, 5000 (1999).



\bibitem{landman} 
C. Yannouleas, E. N. Bogachek, and U. Landman, Phys. 
Rev. B {\bf 57}, 4872 (1998).

\bibitem{lda} 
H. Hakkinen and M. Manninen, Europhys. Lett. {\bf 44}, 80 (1998);
N. Zabala, M. J. Puska, and R. M. Nieminen,
Phys. Rev. Lett. {\bf 80}, 3336 (1998);
A. Nakamura, M. Brandbyge, L. B. Hansen, and K. W. 
Jacobsen, ibid. {\bf 82}, 1538 (1999).



\bibitem{cond.wkb} L. I. Glazman, G. B. Lesovik, D. E. Khmel'nitskii, and
R. I. Shekhter, JETP Lett. {\bf 48}, 238 (1988).




\bibitem{constraint}
It should be emphasized that one should not impose the constraint
$N_- =\mbox{const.}$  This would 
require that the positive background be infinitely
soft, in order to adapt to every mesoscopic variation in the electron
density, which is unrealistic.

\bibitem{lang} N. D. Lang, Solid State Phys. {\bf 28}, 225 (1973). 

\end{thebibliography}
\end{document}